\documentstyle[aps,prd,preprint]{revtex}
\multiply\baselineskip by 2

\def\eV{\rm ~ eV}

\def\Mpc{\rm Mpc}
\def\hmpc{h^{-1}\rm ~Mpc}
\def\tg{T_\gamma}

\def\gsim{\mathrel{\raise.3ex\hbox{$>$}\mkern-14mu
              \lower0.6ex\hbox{$\sim$}}}
\def\lsim{\mathrel{\raise.3ex\hbox{$<$}\mkern-14mu
              \lower0.6ex\hbox{$\sim$}}}

\begin{document}
\preprint{CITA-95-9}
\draft

\title{A Warm-Plus-Hot Dark Matter Universe}

\author{R.A.\ Malaney}
\address{Canadian Institute for Theoretical Astrophysics, University of
Toronto,\\
Toronto, ON Canada M5S 1A7}
\author{G.D.\ Starkman}
\address{Department of Physics, Case Western Reserve Univ., 10900 Euclid Ave,
Cleveland, OH 44106 USA}
\author{L.\ Widrow}
\address{Department of Physics, Queen's University,\\
Kingston, ON Canada K7L 3N6}
\date{\today}
\maketitle
\begin{abstract}

We investigate a new hybrid-model universe containing two types
of dark matter, one ``warm'' and the other ``hot''.
The hot component is an ordinary light neutrino with
mass $\sim 25h^2$~eV while the warm component is a sterile
neutrino with mass $\sim 700h^2$~eV.  The two types of dark matter
arise entirely within the neutrino sector and do not
require separate physical origins.
We calculate the linear transfer functions for a representative
sample of warm-plus-hot models.  The transfer functions,
and results from several observational tests of structure formation,
are compared with those for the cold-plus-hot models that have been
studied extensively in the literature.
On the basis of these  tests, we conclude that warm-plus-hot dark matter
is essentially indistinguishable from cold-plus-hot dark matter,
and therefore provides a viable scenario for large scale structure.
We demonstrate that a neutrino mass matrix can be constructed
which provides the requisite dark matter constituents, while remaining
consistent with all cosmological bounds.

\end{abstract}

\section{Introduction}

Growing evidence against both the cold dark matter (CDM) and hot dark
matter (HDM) cosmological models has prompted researchers to turn
to alternative scenarios.  Mixed dark matter (MDM)
cosmologies [1,2], in which the universe contains an admixture of HDM, CDM,
and ordinary matter, have captured most of the
attention.  By properly tuning the amount of HDM
in the mix ($20\% - 30\%$ in most scenarios) one can find models with a
balance of small ($5\hmpc$) and large ($25-50 \hmpc$) scale power
consistent with current observations [2]  ($h$ is the Hubble
constant today in units of $100~{\rm km\, sec^{-1} Mpc^{-1}}$).  Recent
reports [3] from the Los Alamos National Laboratory suggesting
that $\nu_e$ and/or $\nu_\mu$ have mass in the {\it few}
${\rm eV}$ range (just what the MDM models demand) have heightened
interest in these models.

The essential property of HDM is that relativistic dark matter particles
free-stream out of high-density regions, and therefore
fluctuations are damped
on scales smaller than the horizon size at the epoch when the HDM
becomes non-relativistic.
For ordinary
neutrinos, the canonical HDM candidate, the size of the first objects
to form is $\lambda_{\rm FS}=13\Mpc \left (\Omega_\nu h^2\right )^{-1}$,
where $\Omega_\nu h^2=m_\nu/93\, {\rm eV}$ is the energy density
in dark matter divided by the critical density, and $m_\nu$ is
the neutrino mass.  In MDM models, the cold
component clumps on all scales.  By shifting some of the
mass density from hot to cold matter, one boosts small-scale power
relative to large-scale power.  This is the essence of the MDM models.

The greatest weakness of MDM models is that they require two types of
dark matter with comparable mass densities.  The hot component is usually
taken to be an ordinary neutrino while the cold component is thought to be one
of the standard CDM candidates (e.g., the lightest SUSY particle
or the axion).  A single dark matter
candidate requires ``new physics'' beyond the Standard Model of strong
and electroweak interactions.  At first glance, a second type of dark
matter would require adding  another new  sector to the particle physics
theory, with the unexplained coincidence that the two types of dark matter give
comparable contributions to the total mass density of the universe.
(See, however, ref [1] for an attempt to provide a particle physics
connection between hot neutrinos and cold axions.)

In this paper, we explore the possibility that both the hot and cold
dark matter components of a MDM universe are neutrinos.
We propose that one of the active (weak interactions with
matter) neutrinos is the hot component of the dark matter
while one of the sterile neutrinos (no weak interactions)
is a warm component.  For the sterile neutrinos not to be hot,
their mass to temperature ratio,
$m_s/T_s$, must be greater than that of ordinary HDM.  We therefore require
that the sterile neutrinos decouple from the rest of the
plasma prior to the electroweak (EW) phase transition.  During the
subsequent EW and QCD  phase transitions, the sterile neutrinos cool
relative to the coupled plasma (including the active neutrinos).
$T_s$ is lowered and, assuming a fixed energy density, $m_s$ is higher.
$m_s/T_s$ will be roughly
$15$ times higher than that of a standard  ($\sim 25h^2$~eV) HDM neutrino.
This type of particle is usually referred to as warm dark matter (WDM)
and was first considered in the early 1980's [4].  Large scale
structure in a pure WDM universe with $m/T$ this large
is much like large scale structure in a CDM-dominated universe.
Even if the sterile neutrinos decouple between
the QCD and EW phase transitions ($m_s/T_s$ roughly $8$ times
that of standard HDM neutrinos) they still behave much like
cold dark matter [5].

Our main motivation is to show that two types of dark matter can arise
simultaneously from the neutrino sector and
provide a viable scenario for structure formation.  We
consider it a positive feature of the scenario that the two types of
dark matter have masses which differ by only an order of magnitude or
two.  However, readers familiar with the standard lore of
neutrino mass generation, will recognize our neutrino mass spectrum as
unconventional.  In the usual see-saw mechanism
for generating small masses for active neutrinos,
right-handed fields are typically very heavy due to the presence of
large Majorana mass terms $M_R$ ($\sim10^{3-19}{\rm GeV}$).  The Dirac
mass terms, which couple left and right-handed fields, are assumed to
be comparable to the masses of the associated charged leptons or up-type
quarks, {\it
i.e.} $m_D\sim$~MeV-GeV.  Assuming no left-handed Majorana mass terms, the
physical eigenstates corresponding to the active neutrinos have masses
$\sim (m_D^2/M_R)$, provided $m_D\ll M_R$.  While the see-saw
mechanism may be the most attractive scenario for explaining small
(but non-zero) masses for the active neutrinos, there is no {\it a
priori} reason for all three right-handed fields to be so heavy.  The
models constructed later are counterexamples where the physical
state corresponding to one of the sterile neutrino types ({\it i.e.}
mostly right-handed fields) has mass $m_s\sim 700h^2 \,{\rm eV}$.

\section{Large Scale Structure with Two Types of Neutrino}

Our model universe consists of four components: two types of massive
neutrinos, two massless neutrino species that are treated as a
single component, and a mixture of photons, baryons, and electrons
that is treated as a single component ideal fluid.
We assume primordial perturbations that are adiabatic, Gaussian and
scale-free with a spectrum $P_p(k)=Bk^n$, where $k$ is the wavenumber
of the perturbations measured in units of ${\rm Mpc}^{-1}$.  The COBE
DMR experiment probes energy density perturbations on
very large scales where there is little
modification of the primordial spectrum. It finds $B=8.2\times
10^5h^{-4}{\rm Mpc}^4$ and estimates of $n$ that are consistent
with $n=1$ [6,7].
We use this value of $B$ where normalization of the power spectrum is
required and set $n=1$.  We also set $h=0.5$ and
ignore baryons.  $h=0.5$ is the value used in most
studies of CDM and MDM [2,8] though it may be in conflict with
measurements of the Hubble constant [9].  We emphasize that our main
purpose here is to compare warm-plus-hot models with currently
popular models such as MDM and not to carry out a detailed comparison
with the observations.

Since the perturbations at early times are  small,
one can follow the initial stages of their evolution using linear theory.
In fact, linear theory is adequate for studying structures
on all scales significantly larger than $8\hmpc$.
Quantitative tests on smaller scales probe non-linear structures,
and therefore require N-body simulations.

The active neutrinos decouple from the photon-baryon plasma when they
are still relativistic and when the photon temperature $\tg\simeq
1~{\rm MeV}$.  Their background distribution function is therefore
$f_0(p)=\left ( e^{p/T_\nu}+1\right )^{-1}$ where $T_\nu=\left
(4/11\right )^{1/3}\tg$,~ $p=\left (E_\nu^2-m_\nu^2\right )^{1/2}$,~
and $E_\nu$ is the neutrino energy.
We assume that the background distribution
function for the sterile neutrino species also
has a Fermi-Dirac shape; $f_0(p)~=~\beta\left (e^{p/T_s}+1\right )^{-1}$.
This distribution function is appropriate for a neutrino that
decouples when it is relativistic provided $\beta=1$ and
$T_s=\left (10.75/g_*\right )^{1/3}\left (4/11\right )^{1/3}T_\gamma$,
where $g_*$ is the effective number of relativistic degrees of freedom
in the universe at the time of decoupling.  The masses for the two
neutrinos satisfy the relation
$$
m_\nu ~+~ m_s\left (T_s\over T_\nu\right )^3 ~=~ 93\,  \Omega_\nu h^2 \eV~.
\eqno(1)$$
For a particle that decouples between the EW and QCD phase transitions,
$g_*\simeq 61$, whereas a particle that decouples prior to the EW
phase transition has $g_*\simeq 107$.  We refer to models of these
types as WPH1 and WPH2, respectively.

The above discussion assumes that the sterile neutrinos
were coupled to the baryon-photon plasma at some early epoch.  One
possibility is that coupling occurs at very high energies where
 unknown physics (e.g. from Grand Unified Theories) operates.
Alternatively, oscillations between active and sterile neutrinos may
be responsible for bringing the sterile neutrinos into equilibrium [10,11].
In addition,
Dodelson and Widrow [12] have shown that
under certain assumptions, the momentum space distribution function
for sterile neutrinos produced through oscillations
has a Fermi-Dirac shape with $T_s=\left (4/11\right )^{1/3}T_\gamma$,
but $\beta<1$.  These models lead to exactly the same large
scale structure phenomenology as the early-decoupled particle models [5]
and so our discussion of structure formation applies to both.

It is convenient to define the linear transfer function
\noindent $T(k)\equiv
\left (\delta\rho(k)/\rho\right )/\left
(\delta\rho(k\to\infty)/\rho\right )$ which describes the evolution of
the power spectrum through the linear regime.  The evolved linear
power spectrum is then $P(k)=Bk^nT(k)^2$.  The linear
transfer functions for a variety of models are shown in Figure~1.  As
we can see, the WPH2 models are very close to the MDM models; the WPH1
models have a more distinct shape, though even in this case
the differences are likely to be observationally indistinguishable.  A
more quantitative comparison of the models is made by calculating
various integrals (with appropriate window functions) of the power
spectrum.  In particular, we calculate $\sigma_8$, EP (the excess
power on $25~h^{-1}{\rm Mpc}$), and $\sigma_{0.5}$.  Here, $\sigma_L$ is the
mass excess on $L\hmpc$ defined by the integral
$$
\sigma_L = \left (\int {k^2dk\over 2\pi^2}
P(k)W^2(kL)\right )^{1/2}\eqno(2)
$$
where $W(x)=3\left(\sin{x}-x{\cos{x}}\right )/x^3$ is the top hat window
function.
Fluctuations in the mass
density are related to fluctuations in optically-selected galaxies
through biasing, where $b_o\sigma_L$ gives the fluctuation in
these galaxies on the scale $L\,h^{-1}{\rm Mpc}$.
$b_o$, the optical biasing parameter, can depend on scale, though
in the simplest models it is assumed to be constant.  Davis
and Peebles [13] find that $b_o\sigma_8\simeq 1$ and therefore
$1/\sigma_8$ is a measure of the optical bias.

As mentioned above, CDM falls short because it has too little power
on large scales relative to small scales.
Wright et al. [6]  introduce the quantity
$EP\equiv 3.4{\sigma_{25}/\sigma_{8}}$.
This definition is such that $EP=1$ for standard CDM, whereas
consistency with the APM angular correlation function
 requires $EP=1.30\pm 0.15$.
Finally, $\sigma_{0.5}$ gives a rough (and probably overly
pessimistic) approximation for $1+z_{\rm gf}$, where $z_{\rm gf}$
is the redshift for galaxy formation.  (Alternative methods
for estimating $z_{\rm gf}$ can be found in ref.~[14] and
references therein.)

The results for
$\sigma_8$, EP, and $\sigma_{0.5}$ for a variety of models are given in
Table 1.  For MDM, CDM, and HDM, we use analytic fitting functions
for $T(k)$ taken from Holtzman [15].
We see that the WPH1, WPH2, and MDM models with a 30\% hot component
give similar results for EP and $\sigma_8$.
It has been argued [2] that MDM~30\% satisfies
large scale structure tests on $8-25 h^{-1} {\rm Mpc}$ scales,
and therefore these conclusions should hold for our models as well.
We do note, however,
the results for $\sigma_{0.5}$ in Table 1, and also discussion in the
literature (e.g. ref [2]),
suggest that any model with a significant hot component may have trouble
if observations push the redshift of galaxy formation $z_{\rm gf}> 5$.
This situation can be improved by decreasing the amount of hot matter
in the mix, though this also leads to a decrease in
power on large ($25 \hmpc$) scales.  Since galaxy formation depends on
nonlinear and nongravitational physics (e.g., hydrodynamics) the
above conclusions should be used with caution.

\begin{tabular}{|| c | c | c | c | c | c ||}\hline
{}~~Model~~~~ & ~~~~~$m_\nu$~~~~~ & ~~~~~$m_s$~~~~~&
  ~~~~~$\sigma_8$~~~~~ & ~~~~~EP~~~~&
{}~~~  $\sigma_{0.5}$~~~~~	\\ \hline
   HDM     &   23  & ~~ ---~~ &0.87 & 1.46 & 1.0  \\ \hline
 WPH1 20\% &  4.6 & ~~ 100 ~~& 1.00 & 1.18 & 2.1\\ \hline
 WPH1 30\% &  7.0 & ~~ 90 ~~ & 0.93 & 1.27 & 1.6\\ \hline
 WPH2 20\% &  4.6 & ~~ 190~~ & 1.02 & 1.17 & 2.6\\ \hline
 WPH2 30\% &  7.0 &  ~~160~~ & 0.94 & 1.26 & 2.0\\ \hline
 MDM  30\% &  7.0 &  $\gg 10^3$~~ & 0.96 & 1.30 & 2.6 \\ \hline
   CDM     & --- & $\gg 10^3$ ~~& 1.48 & 0.94 & 9.6\\ \hline
\end{tabular}

\bigskip

\noindent{\bf Table 1} Results for $\sigma_8$, EP, and $\sigma_{0.5}$ for a
variety
of models. The results are based on linear theory calculations. Percentages
in column 1 for the hybrid models indicate fraction of dark matter in the mix.

\bigskip

\section{Neutrino Masses -- Models and Constraints}

At temperatures comparable to $T_{\rm QCD}$, sterile neutrinos interact with
ordinary matter by mixing with active neutrinos.
 From detailed studies of neutrino oscillations in the early universe
[10,11] the conditions that the sterile
neutrinos are out of equilibrium below $T_{\rm QCD}$
(and therefore do not spoil the predictions of Standard Big Bang
Nucleosynthesis
[16]) is
$$\delta m^2 < 3.6 \times 10^{-4} (\sin^22\theta_o)^{-2} \eV^2\ \ \ ;
\ \ {\rm for}\ \sin \theta_o \gsim 10^{-3} \ \ \ ,
 \eqno (3)$$
where $\delta m^2= m_s^2-m_a^2$ is the mass squared difference
 between the states,
and $\theta_o$ is the vacuum mixing angle between them.
This is provided that the energy of the states $E \gg \sqrt{\delta m^2}$,
so that they do not decohere.
However, for WPH1
it does not matter if oscillations bring the sterile neutrinos
into equilibrium above $T_{\rm QCD}$.
As long as the sterile neutrinos decouple prior to the QCD phase transition,
entropy transferred from the quark-gluon plasma
into the interacting gas
will dilute the sterile neutrino number density relative to the
active neutrino number density. For this reason there is no constraint
for $\sin \theta_o \lsim 10^{-3}$
(The exact limit depends on the adopted value for $T_{\rm QCD}$).
For WPH2, the demand that the sterile neutrinos  are out of equilibrium
below the EW phase transition at $T_{\rm EW}\sim 300$~GeV, is essentially that
of Eq.~(3) except that the region where no constraint applies becomes
$\sin \theta_o\lsim 10^{-8}$.
(However, these bounds are model dependent.
In the singlet majoron model [17],
for example, they are greatly weakened
due to the contribution of the majoron background
to the active-neutrino self-energy [18].)

A specific mass matrix which is useful for further discussion is

$$
\pmatrix{0&0&0&0&m_{e\beta}&m_{e\gamma}\cr
0&0&0&0&m_{\mu\beta}&m_{\mu\gamma}\cr
0&0&0&0&m_{\tau\beta}&m_{\tau\gamma}\cr
0&0&0&M'&0&0\cr
m_{e\beta}&m_{\mu\beta}&m_{\tau\beta}&0&M_{\beta\beta}&M_{\beta\gamma}\cr
m_{e\gamma}&m_{\mu\gamma}&m_{\tau\gamma}&0&M_{\beta\gamma}&M_{\gamma\gamma}\cr}\>, \eqno(4)
$$
where we label our weak  eigenstates as
$$
\nu_w=\biggl ({\nu_{e(L)}~ \nu_{\mu (L)}
{}~\nu_{\tau(L) }
{}~ \nu_{\alpha(R)} ~\nu_{\beta(R)} ~\nu_{\gamma(R)}\biggr )
}\> .
 \eqno(5)
$$
There is no reason to identify $\alpha$, $\beta$ and $\gamma$,
with $e$, $\mu$ and $\tau$.
$M_{\beta\gamma}$ is kept non-zero only to simplify later calculations.
In (4) the matrix elements $m_{ij}$ and $M_{ij}$
correspond to Dirac  and Majorana mass terms, respectively.
As in most models, the upper left $3\times 3$ submatrix of (4) is taken
to be zero to be consistent with measurements of the width of the $Z_0$
(see however [19]).
The  other
vanishing terms of  (4) are set to zero for clarity,
and in general they need not be identically zero
for our scenario to remain viable (see later discussion).

Diagonalization of (4) leads to  six mass eigenstates,
which in the limit $m_{ij}\sim m\ll M_{ij}\sim M$ have mass eigenvalues
$$m_1=0; \  m_2,m_3 = {\cal O}\left({m^2\over M}\right);
 \ m_4= M'; \ m_5,m_6\simeq M   \ \ . \eqno(6)$$
 For judicious choices of the entries in (4),
the physical state corresponding to either $m_2$ or $m_3$
will correspond to the (active) HDM component and
that corresponding to $m_4$
will be the (sterile) WDM component.
Those corresponding to $m_5$ and $m_6$
can be made sufficiently heavy with decay rates
(into some combination of light neutrinos and possibly light scalars)
short compared to the age of the Universe.
The mass eigenstates corresponding to $m_2$ and $m_3$
will be a mixture of the $\nu_{e(L)}$, $\nu_{\mu(L)}$ and $\nu_{\tau(L)}$
states
but with a small admixture ${\cal O}(m/M)$ of
the sterile $\nu_{\beta(R)}$ and $\nu_{\gamma(R)}$.
If $m/M$ were large enough, we see from (4) that
the constraint
${m^2/M}< 5 \times 10^{-3}\ {\rm eV} $
would be imposed.
Since $m^2/M$  is the mass of our light active state,
this would imply that  our HDM candidate
could not possess a cosmologically interesting mass.
However, if we take $M \sim 1 $~TeV, then $E<\sqrt{(\delta m^2)}$. As such,
no oscillations will occur and we are relieved of this mass limit.
The mixing angle with the lighter sterile state $\nu_4$ is zero
so there is no oscillations into this state.
 From this we see  that
a mass matrix with $m\sim 3$~MeV, $M'\sim 200\, {\rm eV}$ and $M\sim 1$~TeV,
would satisfy the oscillation constraints and yield the desired neutrino
mass spectrum.

The mass matrix (4) has been chosen to elucidate our discussion.
The first key feature of our scenario is the introduction of
a new mass scale $M'\ll M$.
$M$ may be the vacuum expectation value of one Higgs field
and $M'$ that of another.
Alternatively,
there may be only one new Higgs with different
Yukawa couplings to the right-handed neutrino states.
The smallness of $M'/M$ is the main
\lq\lq{unnatural}\rq\rq~ aspect of our scenario.
However, to put it into perspective,
the above-quoted value for $M'/M \simeq 10^{-9}$ is not enormously smaller
than some of Yukawa couplings in the Standard Model,
such as $g_Y(e) = m_e/M_{weak} \simeq 10^{-6}$,
and is much larger than the ratio of Higgs vacuum expectation values
postulated in Grand Unified Theories, $M_W/M_{GUT}\simeq 10^{-14}$.

A second  feature  of our scenario is the non-participation
(or at least reduced participation)
of the light sterile state in
the usual mixings of the left and right-handed neutrinos
and in the consequent see-saw mechanism.
In (4) this is accomplished by the introduction of
matrix terms identically equal to zero.
These zeros may be enforced by the imposition of global U(1) symmetries.
 For example, in the singlet majoron model [17]
the Dirac mass terms arise from
the coupling of the neutrinos to the
the usual Standard Model Higgs doublet, $H$.
In addition, one introduces a complex scalar (Higgs) field $\Phi$
which is a singlet under SU(2)xU(1) gauge transformations,
but carries non-zero Lepton-number.
$\Phi$ gets a vacuum expectation value generally taken to be
$\langle\Phi\rangle \approx {\rm GeV}-10\,{\rm TeV}$.
The right-handed Majorana masses come from their coupling to
$\langle\Phi\rangle$.
If one extends this model
by adding two new scalars $\Phi_1$ and $\Phi_2$ (or by adding another
sterile neutrino),
one can easily enforce the explicit zeros of (4) without generating
any others -- for example using the Lepton number assignment:

\begin{tabular}{*{10}{p{.4in}}}
&$\nu_{e(L)}$&$\nu_{\mu(L)}$&$\nu_{\tau(L)}$&
         $\nu_{\alpha(R)}$&$\nu_{\beta(R)}$&$\nu_{\gamma(R)}$&$\Phi_1$&$\Phi_2
$&$ H$ \\
L&$1$&$1$&$1$&$2$&$1$&$1$&$-2$&$-4$&$0$\\
\end{tabular}

In general the zeros of (4) need not be identically zero.
That they be small relative to the other scales in the matrix would,
in most cases, suffice. (The exact value required depends on the details
of the model.) If indeed they are non-zero, then
neutrino oscillations could be the mechanism that brings
the sterile neutrinos into equilibrium at early times.
There are two constraints we must impose on the non-zero mass terms.
 First, the admixture of active states in the heavier sterile
mass eigenstate  must be small enough that oscillations
will not keep the sterile state in equilibrium below
either $T_{\rm EW}$  (for WPH2) or $T_{\rm QCD}$ (for WPH1).
(An explicit calculation in the context of the singlet majoron model
shows that $\nu_4$ is decoupled prior to $T_{\rm EW}$.)
Second, experimental limits on neutrino-less $\beta\beta$-decay require
that [20]
$$\langle m_{\nu_e}\rangle ~=~
 \vert \Sigma_i U^2_{ei} m_{\nu_i} \vert^2 < 1.1\eV,$$
where $U_{ij}$ is the unitary matrix diagonalizing (4).
This imposes limits on the values of $m_{ij}$.

The discussion above provides an existence proof of our scenario.
The mass matrix introduced and the constraints imposed on it
are independent of the underlying particle physics model.
However, there are other conditions we must satisfy,
and to investigate them in any meaningful way
requires the specification of a particular particle physics model.
The model-dependent constraints are:

\noindent
(i) The  decay of the light sterile state $\nu_s$ must be inhibited.
The most general neutrino decay scheme can be written
$\nu_s \rightarrow \nu_a + X$
where $\nu_a$ is some lighter neutrino (in our scenario one of
the active states), and $X$ is some set of
bosons and/or pairs of lighter fermions.

\noindent
(ii) Any new light degrees of freedom (e.g., the majoron)
must be decoupled prior to $T_{\rm QCD}$,
so as not to contribute significantly
to the energy density of the universe at the epoch of nucleosynthesis.
To ensure this, we assume that our new particle does not couple
in any significant way with the active neutrino states or with
ordinary matter. If it couples  strongly only
with the sterile neutrinos, then it to will be diluted away by entropy transfer
at the electroweak and QCD phase transitions.
Again, in the context of the singlet majoron model, this
proves to be the case for the values of $m$, $M'$ and $M$ quoted.

\noindent
(iii)  The heavy degrees of freedom  (here $\nu_{5,6}$)
must annihilate or decay away before they
can dominate the energy density of the universe.
Again, in the  singlet majoron model, the heavy states decay away
on a sufficiently short time scale.

\noindent
(iv) new $\beta\beta$-decay modes must be adequately suppressed.
In the singlet majoron model, $g_{eeJ}$ the effective  coupling of the Majoron
to
$\nu_e$ must be $\leq 7\times 10^{-5}$ [21].  This is satisfied by the
model presented here.

The above constraints are clearly model dependent in that they
rely on the details of the theory which
ascribes mass to the neutrino sector.
Any detailed investigation of the validity of our scenario
therefore requires knowledge of the underlying  particle physics.
We have  sometimes highlighted the possibilities for our scenario
in the context of the singlet majoron model.
 Many other models possessing sterile neutrino fields,
including more complicated majoron models,
can also be constructed.

\section{Conclusion}

We have investigated the possibility that the universe possesses
a mixture of warm and hot dark matter.
The hot component is identified with a known light neutrino
with mass $\sim 25 h^2$~eV; and
 the warm component with a sterile neutrino
with mass $\sim 700 h^2$~eV.  These mass ranges satisfy
the normal cosmological mass limits for stable neutrinos, provided
the sterile neutrinos decouple at an early epoch. From calculations
of the linear transfer functions for such a hybrid model, we show
that mixed warm-plus-hot dark matter  is a viable cosmological
scenario, and in most respects
provides a better fit to observational data than either the standard HDM or CDM
models. It is also interesting to note that the $700h^2$~eV sterile neutrino
can evade the usual phase space constraints [22].

The introduction of our warm dark matter particle has come at a price,
namely, the neglect of  theoretical prejudice that all right-handed
neutrino fields be very massive ($m>$GeV).
We emphasize, however,
that {\it a priori} the mass terms  of matrix (4) are unknown, and
there is no fundamental reason that sterile neutrinos
cannot possess masses in the $700h^2~{\rm eV}$ range.
In addition,
 we have discussed how
such particles satisfy  all standard cosmological bounds
 given certain restrictions
on their oscillation  properties.
If theoretical prejudice is relaxed just a little,
we see that a solution to the problems of
the formation of large scale structure
could reside purely in the neutrino sector.

We find it suggestive
that one can accommodate a hybrid dark matter universe
entirely within one sector avoiding the need for separate
physical origins of the dark matter constituents.
Given these arguments, and given  the recent tentative
experimental evidence for a hot neutrino component, we believe
this model deserves further consideration.

\section*{Acknowledgements}

We thank J. R. Bond, M. Butler, S. Dodelson, and
L. Krauss for useful discussions.
GDS acknowledges support for this research from the Canadian Institute for
Advanced Research, Cosmology Program and the Canadian Institute for
Theoretical Astrophysics (C.I.T.A.) where this work was begun, and from
the College of Arts and Sciences, C.W.R.U.  GDS also thanks C.I.T.A.
for continued hospitality during the course of this research.
LMW acknowledges support from the Natural Sciences and Engineering
Research Council of Canada.

\noindent


\begin{references}


\bibitem{}~Q.~Shafi and F.~W.~Stecker, Phys.~Rev.~Lett. {\bf 53}, 1292 (1984).

\bibitem{}~M.~Davis, F.~Summers, and D.~Schlegel, Nature  {\bf 359}, 393
(1992);
A.~N.~Taylor and M.~Rowan-Robinson, Nature, {\bf 359}, 396 (1992);
A.~van Dalen and R.~K.~Schaefer, Astrophys.~J. {\bf 398}, 33 (1992);
A.~Klypin, J.~Holtzman, J.~ Primack, and E.~ Reg\"os, Astrophys.~J.~,
{\bf 416}, 1 (1993).

\bibitem{}New York Times, Jan. 31, 1995; and J. Margulies,
talk at "Unified Symmetry in the Small and in the Large,"
Miami, FL, Feb. 2-5, 1995.

\noindent
\bibitem{}J.~R.~Bond, A.~S.~Szalay, and M.~S.~ Turner, Phys.~ Rev.~ Lett.
{\bf  48}, 1636 (1982);
J.~R.~Bond, and A.~S.~ Szalay, Astrophys.~J.~{\bf  274}, 443 (1983).

\noindent
\bibitem{}S.~Colombi, S.~Dodelson, and L.~M.~Widrow, {\it in preparation}
(1995).

\noindent
\bibitem{}E.~L.~Wright, et al.~, Astrophys.~J.~ {\bf396}, L13 (1992).

\noindent
\bibitem{}G.~Efstathiou, J.~R.~Bond, and S.~D.~M.~White,
Mon.~Not.~R.~astr.~Soc.~,
{\bf 258}, 1p (1992); C.~L.~Bennett et al., NASA/Goddard preprint (1995).

\noindent
\bibitem{}M.~Davis, G.~Efstathiou, C.~S.~ Frenk, and
S.~D.~M.~ White,  Astrophys.~J.~, {\bf 292}, 371 (1985) and references
therein.

\noindent
\bibitem{}M.~J.~Pierce et al., Nature {\bf 371}, 385 (1994);
W.~L.~Freedman, et al.~, Nature {\bf 371}, 757 (1994) and references therein.

\noindent
\bibitem{}R. Barbieri and A. Dolgov, Phys. Letts.  {\bf  237B}, 440 (1990);
 R. Barbieri and A. Dolgov, Nucl. Phys {\bf 349B}, 743 (1991).

\noindent
\bibitem{}K. Kainailunen, Phys. Lett. {\bf 244B}, 191 (1990).

\noindent
\bibitem{}S.~ Dodelson and L.~M.~Widrow, Phys.~Rev.~Lett.~{\bf 72}, 17 (1994).

\noindent
\bibitem{}M.~Davis, and P.~J.~E.~Peebles, Astrophys.~J.~{\bf 267}, 465 (1983).

\noindent
\bibitem{}F.~C.~Adams, et al.~, Phys.~Rev.~{\bf D47}, 426 (1993).

\noindent
\bibitem{}J.~A.~Holtzman, Astrophys.~J.~Supp., {\bf 71}, 1 (1989).

\noindent
\bibitem{}L.~M.~Krauss and P.~J.~Kernan
CWRU-P9-94,  Phys. Lett. B.  in press (1995), and references therein.

\noindent
\bibitem{}Y.~Chikashige, R.~N.~Mohapatra and R.~D.~Peccei,
Phys. Lett {\bf 98B}, 265 (1981).

\noindent
\bibitem{}K.~S.~Babu and I.~Z.~Rothstein, Phys. Lett. {\bf  275B}, 112 (1992).

\noindent
\bibitem{}K.~Choi and A.~Santamaria, Phys. Lett. {\bf  267B}, 504 (1991).

\noindent
\bibitem{}B.~Maier, Nucl. Phys (Proc. Suppl.) B35, 358 (1994)

\noindent
\bibitem{}M.~K.~Moe, Nucl. Phys (Proc. Suppl.) B35, 386 (1994)

\noindent
\bibitem{}S.~Tremaine and J.~E.~Gunn, Phys. Rev. Lett. {\bf 42}, 407 (1979).




\end{references}
\end{document}